\documentclass[reprint,
superscriptaddress,
amsmath,amssymb,aps,showkeys,showpacs,
twoside,final,secnumarabic,
nofootinbib]{revtex4-2}

\usepackage[paperwidth=205mm,paperheight=290mm,top=17mm,bottom=25mm,
inner=17mm,outer=17mm,
twoside]{geometry}

\usepackage{cmap} 
\usepackage[T1,T2A]{fontenc}
\usepackage[utf8]{inputenc}
\usepackage[russian,english]{babel}
\usepackage{color}
\usepackage{graphicx}
\usepackage{dcolumn}
\usepackage{bm} 
\usepackage[unicode=true,colorlinks=true,linkcolor=magenta, urlcolor=blue, citecolor = blue,breaklinks]{hyperref}
\usepackage{multirow}
\usepackage{url}
\usepackage{breakurl}
\DeclareGraphicsExtensions{.eps}

\newcount\issue
\newcount\Vol
\newcount\numb
\usepackage{fancyhdr} 
\def\Vol{\textbf{78}}
\def\numb{x}
\begin{document}

\title{CONFERENCE SECTION\\[20pt]
A Relativistic MOND} 

\def\addressa{Tata Institute of Fundamental Research, Mumbai 400005, India}

\author{\firstname{Tejinder P. }~\surname{Singh}}
\email[E-mail: ]{tpsingh@tifr.res.in }
\affiliation{\addressa}

\received{xx.xx.2026}
\revised{xx.xx.2026}
\accepted{xx.xx.2026}

\begin{abstract}
\noindent We present a minimal relativistic completion of MOND in which (i) General Relativity is recovered
exactly in the high-acceleration regime, while (ii) the Bekenstein--Milgrom (AQUAL) equation
emerges in the low-acceleration regime, without introducing additional propagating fields beyond
those already present in a right-handed gauge sector.
The construction is motivated by an $E_6\times E_6$ framework in which
$SU(3)_R\rightarrow SU(2)_R\times U(1)_{Y'}\rightarrow U(1)_{\rm dem}$, leaving a healthy repulsive
$U(1)_{\rm dem}$ interaction whose charge is the square-root mass label.
Gravity itself arises from the $SU(2)_R$ connection via a Plebanski/MacDowell--Mansouri mechanism,
yielding an emergent tetrad and the Einstein--Hilbert action.
MOND is implemented by an infrared (IR) metric deformation $\Delta S_{\rm IR}[g]$ that is
UV-vanishing (so GR is recovered) while its deep-MOND/static limit is fixed by a symmetry principle:
in three spatial dimensions, the deep-MOND action is conformally invariant with a 10-parameter group
isomorphic to $SO(4,1)$ (the de Sitter group).
The single MOND acceleration scale is set by a de Sitter radius selected dynamically in the IR,
$a_0=c^2/(\xi\,\ell_{\rm dS})$ with $\xi=\mathcal{O}(1)$ fixed by matching to the static limit.
MOND resides in perturbations and quasistatic systems; the homogeneous FRW background is controlled
by the IR vacuum kinematics rather than an ad hoc cosmological constant.
\end{abstract}

\par
\keywords{MOND; deSitter vacuum; Relativistic MOND; $SU(2)_R$ gauge symmetry    \\[5pt]}

\maketitle
\thispagestyle{fancy}

\vspace{-0.3cm}
\section{Motivation: a minimal relativistic MOND}
Empirically, galaxy dynamics exhibit a low-acceleration regularity encapsulated by MOND \cite{Milgrom1983, BekensteinMilgrom1984}:
when characteristic accelerations fall below a universal scale $a_0\sim 10^{-10}\,{\rm m\,s^{-2}}$,
the relation between baryonic mass and asymptotic rotation velocity becomes
$v^4\simeq G a_0 M_b$ (the baryonic Tully--Fisher relation), and rotation curves correlate tightly
with baryonic distributions (the radial acceleration relation).
A relativistic completion should: (i) reduce to GR at high acceleration (Solar System), (ii) reproduce
MOND in the deep IR, (iii) yield correct lensing, and (iv) remain ``healthy'' (no ghosts, no wrong-sign
kinetic terms). Many relativistic MOND theories introduce additional scalar/vector degrees of freedom.
Here the goal is more restrictive: obtain MOND through a \emph{metric-only} IR deformation whose form
and scale are selected by an IR vacuum principle plus a deep-IR symmetry.

\vspace{-0.2cm}
\section{Right-handed sector and a healthy \texorpdfstring{$U(1)_{\rm dem}$}{U(1)dem}}
We work within a schematic $E_6\times E_6$ setting \cite{SinghUni} and focus on the right-handed breaking chain
\begin{equation}
SU(3)_R \;\rightarrow\; SU(2)_R\times U(1)_{Y'} \;\rightarrow\; U(1)_{\rm dem}.
\label{eq:breaking}
\end{equation}
The unbroken Abelian generator is taken to be a square-root mass label with eigenvalues
$s=\pm\sqrt{m/\kappa}$. The corresponding vector interaction is standard and repulsive for like signs,
\begin{equation}
\mathcal{L}_{\rm dem} = -\frac{1}{4}F_{\mu\nu}F^{\mu\nu} + g_{\rm dem} A_\mu J^\mu_s,
\qquad
J^\mu_s=\sum_\psi \bar\psi\gamma^\mu S_{\rm dem}\psi.
\label{eq:dem}
\end{equation}
This $U(1)_{\rm dem}$ field is \emph{not} the mediator of the MOND force: MOND will arise from the
gravitational sector via an IR deformation of the metric action. The role of $U(1)_{\rm dem}$ is
conceptual (as an unbroken remnant of the right-handed sector) and can be arranged to remain
subleading in galactic phenomenology.

\vspace{-0.2cm}
\section{Gauge gravity from \texorpdfstring{$SU(2)_R$}{SU(2)R} and the GR limit}
Let $\omega^i{}_\mu$ be the $SU(2)_R$ connection with curvature $F^i$.
A Plebanski/MacDowell--Mansouri \cite{Plebanski1977, MacDowellMansouri1977} seed action supplemented by algebraic simplicity constraints
generates Einstein gravity after a soldering/transition sector implements the identification with
spacetime geometry:
\begin{equation}
S^{SU(2)_R}_{BF+{\rm cons}}=
\frac{1}{8\pi G}\int B^i\wedge F^i
\;+\;\int \lambda_{ij} B^i\wedge B^j
\;+\; S_{\rm trans}[{\rm Higgs}_R].
\label{eq:bf}
\end{equation}
The non-propagating multipliers $\lambda_{ij}$ enforce the simplicity constraint
$B^i \propto \epsilon^i{}_{jk}\,e^j\wedge e^k$, producing an emergent tetrad $e^I{}_\mu$ and metric
$g_{\mu\nu}=e^I{}_\mu e^J{}_\nu \eta_{IJ}$. One then recovers the Einstein--Hilbert action
\begin{equation}
S_{\rm EH}[g]=\frac{1}{16\pi G}\int d^4x\sqrt{-g}\;R[g].
\label{eq:eh}
\end{equation}
Thus, in the regime where additional IR effects are negligible, the theory reduces to GR without
introducing new gravitational propagating degrees of freedom.

\vspace{-0.2cm}
\section{de Sitter IR vacuum selection and the MOND scale}
The key input is an \emph{IR vacuum principle}: the deep IR of the right-handed sector realizes
de Sitter (dS) kinematics. Let $\ell_{\rm dS}$ be the dS radius selected dynamically by the $SU(2)_R$
vacuum. Dimensional analysis then ties the MOND acceleration scale to this length,
\begin{equation}
a_0=\frac{c^2}{\xi\,\ell_{\rm dS}},
\label{eq:a0}
\end{equation}
where $\xi=\mathcal{O}(1)$ is fixed by matching the relativistic action to the static (AQUAL) limit.
In this sense $a_0$ is not inserted as an independent parameter: it is determined by the IR dS vacuum.

A useful bookkeeping device (especially for interpreting static galactic phenomenology in an expanding
universe) is an ``effective distance'' defined by
\begin{equation}
r_{\rm eff}^2 = R(t)\,R_H(t),\qquad R_H(t)\equiv \frac{a}{\dot a},
\label{eq:reff}
\end{equation}
with $R(t)$ the FRW proper distance and $R_H$ the Hubble radius.
In deep-MOND phenomenology one may freeze $R_H$ to its present value, effectively rendering $a_0$
epoch-independent for late-time galactic dynamics, while the far-IR vacuum remains dS-like.

\vspace{-0.2cm}
\section{Metric-only IR deformation and the AQUAL limit}
We introduce a dimensionless invariant
\begin{equation}
y \equiv \frac{I[g]}{a_0^2},
\qquad
I[g]\equiv a_\mu a^\mu,\qquad a_\mu \equiv \nabla_\mu\ln N,
\label{eq:I}
\end{equation}
with $N$ the lapse (in Newtonian gauge, $N=\sqrt{-g_{00}}$). For $g_{00}=-(1+2\Phi)$ one has
$\ln N\simeq \Phi$, hence $I[g]\rightarrow |\nabla\Phi|^2$ in the static weak-field limit.
We define the \emph{relativistic MOND regime} by $y\ll 1$ (i.e. $\sqrt{a_\mu a^\mu}\ll a_0$), while the
GR regime corresponds to $y\gg 1$.

\subsection*{No double counting and the two regimes}
In the nonrelativistic $00$-sector, $S_{\rm EH}$ already yields the standard Newtonian quadratic piece.
Explicitly,
\begin{align}
S_{\rm EH}\ \longrightarrow\ -\int dt\,\frac{1}{8\pi G}\int d^3x\,|\nabla\Phi|^2\nonumber\\
\;=\;-\int dt\,\frac{a_0^2}{8\pi G}\int d^3x\,y.
\label{eq:EH_NR}
\end{align}
An AQUAL/MOND functional has a GR limit at large $y$ (equivalently $\mu\to 1$), so adding it naively
would double-count the quadratic term in the high-acceleration regime. We therefore work with the
UV-vanishing deformation
\begin{equation}
\Delta S_{\rm IR}[g]\ \equiv\ -\,\frac{a_0^2}{8\pi G}\int d^4x\sqrt{-g}\,
\Big[\,F(y)-y\,\Big].
\label{eq:DeltaSIR}
\end{equation}
For $y\gg 1$, recovery of GR requires $F'(y)\rightarrow 1$ (equivalently $\mu\rightarrow 1)$, so the
contribution of $\Delta S_{IR}$ to the field equations is suppressed in the high acceleration regime. For the choice (14), $F(y)=-2\sqrt{y} +{ \cal O}(\ln y)$ is subleading 
compared to $y$ and the resulting corrections scale as $y^{-1/2}\sim a_0/|\nabla \Phi|$.

In the static limit ($I[g]\to |\nabla\Phi|^2$), the sum $S_{\rm EH}+\Delta S_{\rm IR}$ reduces to the
standard AQUAL functional
\begin{equation}
S_{\rm stat}[\Phi]=-\int dt\,\frac{a_0^2}{8\pi G}\int d^3x\;
F\!\left(\frac{|\nabla\Phi|^2}{a_0^2}\right)\;+\;\int dt\int d^3x\,\rho\,\Phi,
\label{eq:AQUAL_stat}
\end{equation}
whose Euler--Lagrange equation is the Bekenstein--Milgrom equation
\begin{equation}
\nabla\cdot\!\left[\mu\!\left(\frac{|\nabla\Phi|}{a_0}\right)\nabla\Phi\right]=4\pi G\rho,
\qquad
\mu(x)\equiv F'(x^2).
\label{eq:BM}
\end{equation}

\subsection*{Deep-MOND conformal symmetry, $SO(4,1)$, and the de Sitter connection}
Milgrom has emphasized that the deep-MOND limit for purely gravitational nonrelativistic systems can
be characterized by a \emph{spacetime scaling symmetry} of the equations of motion,
$(t,\mathbf{r})\to(\lambda t,\lambda \mathbf{r})$ in the formal limit $a_0\to\infty$ \cite{Milgrom2009}.
For a single-potential, action-based ``modified gravity'' formulation, this selects (up to normalization)
the deep-MOND Lagrangian density $\propto |\nabla\Phi|^3/a_0$, i.e.
\begin{equation}
\mathcal{L}_{\rm deep}\ \propto\ \frac{|\nabla\Phi|^3}{a_0}
\qquad\Longleftrightarrow\qquad
F(y)\ \propto\ y^{3/2}\quad (y\ll 1),
\label{eq:deep}
\end{equation}
since $y=|\nabla\Phi|^2/a_0^2$ in the static limit.
In $d=3$ spatial dimensions, the resulting deep-MOND field equation also enjoys invariance under the
full 10-parameter conformal group of Euclidean space (3 translations, 3 rotations, 1 dilation, and
3 special conformal transformations) \cite{Milgrom2009}. This group is isomorphic to $SO(4,1)$,
which is also the isometry group of $dS_4$.

In our framework, the assumption that the right-handed $SU(2)_R$ sector flows to a de Sitter IR fixed
point provides both the length scale $\ell_{\rm dS}$ (hence $a_0$ via~\eqref{eq:a0}) and the symmetry
criterion: the deep-MOND/static sector should realize the same $SO(4,1)$ symmetry, manifested as the
3D conformal invariance of~\eqref{eq:deep}. This fixes only the \emph{asymptotic} form $F(y)\sim
\frac{2}{3}y^{3/2}$ as $y\to 0$; the full theory is then obtained by requiring $F'(y)\to 1$ as $y\to\infty$
so that GR is recovered.

A convenient parameter-free choice is the interpolation function
\begin{equation}
\mu(x)=\frac{x}{1+x},\qquad x\equiv \frac{|\nabla\Phi|}{a_0},
\label{eq:mu}
\end{equation}
which corresponds to $F'(y)=\mu(\sqrt{y})$ and hence
\begin{equation}
F(y)=y-2\sqrt{y}+2\ln\!\big(1+\sqrt{y}\big),
\label{eq:F}
\end{equation}
(up to an irrelevant constant). This reproduces the deep-MOND scaling $F(y)\sim \frac{2}{3}y^{3/2}$
as $y\to 0$, and $F'(y)\to 1$ as $y\to\infty$.

\subsection*{Field equations and the static MOND limit}
The microscopic gravitational sector is the $SU(2)_R$ BF+constraints action~\eqref{eq:bf} (which
already includes the soldering/transition term $S_{\rm trans}$). Imposing the simplicity constraint and
integrating out the auxiliary fields yields the metric Einstein--Hilbert action~\eqref{eq:eh}. Hence, for
phenomenology we work at the metric level and do not include $S^{SU(2)_R}_{BF+{\rm cons}}$ and
$S_{\rm EH}$ simultaneously. The effective action is
\begin{align}
S_{\rm total}[g,A,\Psi]\equiv
S_{\rm EH}[g]\;+\;\Delta S_{\rm IR}[g]\;+\;S_{\rm dem}[A]\;+\nonumber\\ \;S_{\rm matter}[g,\Psi].
\label{eq:Stotal}
\end{align}
Varying with respect to $g_{\mu\nu}$ yields modified Einstein equations
\begin{equation}
G_{\mu\nu}+\Xi_{\mu\nu}[g;a_0,F]=8\pi G\left(T^{\rm matter}_{\mu\nu}+T^{\rm dem}_{\mu\nu}\right),
\label{eq:Einstein_mod}
\end{equation}
where $\Xi_{\mu\nu}\equiv -(2/\sqrt{-g})\,\delta \Delta S_{\rm IR}/\delta g^{\mu\nu}$.
In the static weak-field limit, variation with respect to $\Phi$ gives~\eqref{eq:BM}.
The two limits follow immediately:
\begin{align}
|\nabla\Phi|\gg a_0: &\quad \mu\to 1\;\Rightarrow\;\nabla^2\Phi=4\pi G\rho,\\
|\nabla\Phi|\ll a_0: &\quad \mu(x)\sim x\;\Rightarrow\;
\nabla\cdot\left(|\nabla\Phi|\nabla\Phi\right)=4\pi G a_0\rho\; \nonumber\\\Rightarrow
\; v^4=G a_0 M.
\end{align}
By construction, in the quasistatic regime there is no gravitational slip ($\Psi=\Phi$), so lensing is
governed by the same potential that controls dynamics.

\vspace{-0.2cm}
\section{Cosmological remarks and the MOND--de Sitter connection}
On an FRW background written in cosmic time, one has $N=1$ and hence $a_\mu=0$, so $I[g]=0$ and
$\Delta S_{\rm IR}$ does not modify the homogeneous background equations. The far-IR vacuum
nonetheless selects a dS kinematics with an effective curvature scale
$\Lambda_{\rm eff}\sim 3c^2/\ell_{\rm dS}^2$, fixed by the right-handed IR vacuum rather than inserted
as a free cosmological constant.

Milgrom has stressed two related facts \cite{Milgrom2009}: (i) the empirical proximity
$\bar a_0\equiv 2\pi a_0 \sim cH_0 \sim c^2/\ell_{\rm dS}$ (``cosmic coincidence''), and (ii) the
equivalence between the $dS_4$ isometry group $SO(4,1)$ and the 10-parameter conformal group acting
on three-dimensional Euclidean space. He conjectures that in an exact de Sitter universe local gravity
might approach the \emph{deep-MOND} form, and notes possible relevance of a dS/CFT perspective.

Our approach differs in emphasis. We \emph{postulate} a right-handed $SU(2)_R$ IR vacuum that is
de Sitter and thereby \emph{derives} a preferred length $\ell_{\rm dS}$, which sets $a_0$ via
\eqref{eq:a0}; the deep-MOND $SO(4,1)$ symmetry is then implemented directly at the level of the
static IR functional through the asymptotic condition~\eqref{eq:deep}. In particular, we do not require
that an exact dS cosmology forces \emph{all} local systems into the deep-MOND regime; rather, deep MOND
still corresponds to the local invariant threshold $y\ll 1$. Identifying $\ell_{\rm dS}$ with the asymptotic
cosmological dS radius would make $\xi$ in~\eqref{eq:a0} numerically comparable to Milgrom's
$2\pi$ factor.

This structure yields clear observational handles once $a_0$ is fixed:
(i) baryonic Tully--Fisher and the radial acceleration relation with small intrinsic scatter, since
$a_0$ is tied to a cosmological scale;
(ii) enhanced late-time structure growth when the effective gravitational response is boosted in the
low-acceleration regime;
(iii) lensing without slip, hence predictable correlations between dynamical and lensing masses across
the GR--MOND crossover;
(iv) late-time ISW and CMB lensing modifications arising from the time evolution of $\Phi$ induced
by the MOND closure.

\vspace{-0.2cm}
\section{Discussion}
A relativistic MOND can be achieved with a metric-only, UV-vanishing IR deformation:
GR is recovered exactly at high acceleration, while AQUAL/MOND emerges at low acceleration.
The deep-MOND/static sector is selected by a symmetry principle---3D conformal invariance with
group $SO(4,1)$---which is naturally suggestive of an underlying de Sitter IR fixed point. In the present
construction the dS radius is supplied by the right-handed $SU(2)_R$ vacuum and sets the MOND
acceleration scale via~\eqref{eq:a0}.
A central open task is to promote the present ``cosmological-rest-frame'' implementation of $I[g]$ into
a fully covariant completion (or to show it is sufficient), and to develop the cosmological perturbation
theory in detail.

\noindent{\bf Covariant completion and diffeomorphism invariance.}
The definition $a_\mu\equiv\nabla_\mu\ln N$ in Eq.~(7) uses the lapse $N$ (in practice
$N=\sqrt{-g_{00}}$ in Newtonian gauge), and is therefore simplest in a preferred foliation
(the cosmological rest frame) rather than manifest 4D diffeomorphism invariance.
A fully covariant completion can be pursued in three logically distinct ways:
(A) introduce a unit timelike field $u^\mu$ (or a scalar ``clock'' $T$ with
$u_\mu\propto\nabla_\mu T$) and replace $a_\mu$ by the covariant 4-acceleration
$a_\mu=u^\nu\nabla_\nu u_\mu$;
(B) identify $u^\mu$ with the matter rest-frame (e.g.\ the baryonic 4-velocity) wherever this is
well defined; or
(C) keep the theory metric-only but define the foliation as a covariant functional of $g_{\mu\nu}$
(e.g.\ constant-mean-curvature slicing), which is typically nonlocal.
In this proceedings we treat the cosmological-rest-frame implementation as an effective description;
determining which of (A--C) is realized by the underlying $SU(2)_R$ vacuum, and the resulting
implications for perturbations and Lorentz/diffeomorphism tests, is left for future work.


\vspace{-0.2cm}
\begin{thebibliography}{9}
\bibitem{Milgrom1983} M. Milgrom, \emph{A modification of the Newtonian dynamics as a possible alternative to the hidden mass hypothesis}, Astrophys. J. \textbf{270} (1983) 365.
\bibitem{BekensteinMilgrom1984} J. Bekenstein and M. Milgrom, \emph{Does the missing mass problem signal the breakdown of Newtonian gravity?}, Astrophys. J. \textbf{286} (1984) 7.
\bibitem{SinghUni} T. P. Singh, \emph{Trace dynamics, octonions, and unification} J. Phys. Conf. Ser. 2912, 012009 (2024) arXiv:2501.18139
\bibitem{Plebanski1977} J. F. Plebanski, \emph{On the separation of Einsteinian substructures}, J. Math. Phys. \textbf{18} (1977) 2511.
\bibitem{MacDowellMansouri1977} S. W. MacDowell and F. Mansouri, \emph{Unified geometric theory of gravity and supergravity}, Phys. Rev. Lett. \textbf{38} (1977) 739.
\bibitem{Milgrom2009} M. Milgrom, \emph{The MOND limit from spacetime scale invariance}, Astrophys. J. \textbf{698} (2009) 1630--1638.

\end{thebibliography}
\end{document}